

\font\titlefont = cmr10 scaled\magstep 4
 2
\font\sectionfont = cmr10
\font\littlefont = cmr5
\font\eightrm = cmr8 

\def\ss{\scriptstyle} 
\def\sss{\scriptscriptstyle} 

\newcount\tcflag
\tcflag = 0  

\ifnum\tcflag = 0 \magnification = 1200 \fi  

\global\baselineskip = 1.2\baselineskip
\global\parskip = 4pt plus 0.3pt
\global\abovedisplayskip = 18pt plus3pt minus9pt
\global\belowdisplayskip = 18pt plus3pt minus9pt
\global\abovedisplayshortskip = 6pt plus3pt
\global\belowdisplayshortskip = 6pt plus3pt

\def\barsoff{\overfullrule=0pt}


\def\endignore{}
\def\ignore #1\endignore{}

\newcount\dflag
\dflag = 0


\def\monthname{\ifcase\month
\or January \or February \or March \or April \or May \or June%
\or July \or August \or September \or October \or November %
\or December
\fi}

\newcount\dummy
\newcount\minute  
\newcount\hour
\newcount\localtime
\newcount\localday
\localtime = \time
\localday = \day

\def\advanceclock#1#2{ 
\dummy = #1
\multiply\dummy by 60
\advance\dummy by #2
\advance\localtime by \dummy
\ifnum\localtime > 1440 
\advance\localtime by -1440
\advance\localday by 1
\fi}

\def\settime{{\dummy = \localtime%
\divide\dummy by 60%
\hour = \dummy
\minute = \localtime%
\multiply\dummy by 60%
\advance\minute by -\dummy
\ifnum\minute < 10 
\xdef\spacer{0} 
\else \xdef\spacer{} 
\fi %
\ifnum\hour < 12 
\xdef\ampm{a.m.} 
\else 
\xdef\ampm{p.m.} 
\advance\hour by -12 %
\fi %
\ifnum\hour = 0 \hour = 12 \fi
\xdef\timestring{\number\hour : \spacer \number\minute%
\thinspace \ampm}}}



\def\endtitle{}
\def\title#1\endtitle{\vskip.5in\titlefont
\global\baselineskip = 2\baselineskip
#1\vskip.4in
\baselineskip = 0.5\baselineskip\rm}
 
\def\endauthors{}
\def\authors#1\endauthors{#1}

\def\endabstract{}
\def\abstract#1\endabstract{\vskip .3in%
\centerline{\sectionfont\bf Abstract}%
\vskip .1in
\noindent#1}

\def\nopageonenumber{\footline={\ifnum\pageno<2\hfil\else
\hss\tenrm\folio\hss\fi}}  

\newcount\nsection
\newcount\nsubsection

\def\section#1{\global\advance\nsection by 1
\nsubsection=0
\bigskip\noindent\centerline{\sectionfont \bf \number\nsection.\ #1}
\bigskip\rm\nobreak}

\def\subsection#1{\global\advance\nsubsection by 1
\bigskip\noindent\sectionfont \sl \number\nsection.\number\nsubsection)\
#1\bigskip\rm\nobreak}

\def\topic#1{{\medskip\noindent $\bullet$ \it #1:}} 
\def\endtopic{\medskip}

\def\appendix#1#2{\bigskip\noindent%
\centerline{\sectionfont \bf Appendix #1.\ #2}
\bigskip\rm\nobreak}


\newcount\nref
\global\nref = 1

\def\therefs{} 


\def\ref#1#2{\xdef #1{[\number\nref]}
\ifnum\nref = 1\global\xdef\therefs{\item{[\number\nref]} #2\ }
\else
\global\xdef\oldrefs{\therefs}
\global\xdef\therefs{\oldrefs\vskip.1in\item{[\number\nref]} #2\ }%
\fi%
\global\advance\nref by 1
}

\def\listrefs{\vfill\eject\section{References}\therefs}


\newcount\nfoot
\global\nfoot = 1

\def\foot#1#2{\xdef #1{(\number\nfoot)}
\footnote{${}^{\number\nfoot}$}{\eightrm #2}
\global\advance\nfoot by 1
}


\newcount\nfig
\global\nfig = 1
\def\thefigs{} 

\def\figure#1#2{\xdef #1{(\number\nfig)}
\ifnum\nfig = 1\global\xdef\thefigs{\item{(\number\nfig)} #2\ }
\else
\global\xdef\oldfigs{\thefigs}
\global\xdef\thefigs{\oldfigs\vskip.1in\item{(\number\nfig)} #2\ }%
\fi%
\global\advance\nfig by 1 } 

\def\fig#1{\xdef #1{(\number\nfig)}
\global\advance\nfig by 1 } 


\newcount\ntab
\global\ntab = 1

\def\table#1{\xdef #1{\number\ntab}
\global\advance\ntab by 1 } 


\newcount\cflag
\newcount\nequation
\global\nequation = 1
\def\eqlabel{(1)}

\def\nexteqno{\ifnum\cflag = 0
\global\advance\nequation by 1
\fi
\global\cflag = 0
\xdef\eqlabel{(\number\nequation)}}

\def\lasteqno{\global\advance\nequation by -1
\xdef\eqlabel{(\number\nequation)}}

\def\label#1{\xdef #1{(\number\nequation)}
\ifnum\dflag = 1
{\escapechar = -1
\xdef\draftname{\littlefont\string#1}}
\fi}

\def\clabel#1#2{\xdef\eqlabel{(\number\nequation #2)}
\global\cflag = 1
\xdef #1{\eqlabel}
\ifnum\dflag = 1
{\escapechar = -1
\xdef\draftname{\string#1}}
\fi}

\def\cclabel#1#2{\xdef\eqlabel{#2)}
\global\cflag = 1
\xdef #1{\eqlabel}
\ifnum\dflag = 1
{\escapechar = -1
\xdef\draftname{\string#1}}
\fi}


\def\eeq{}

\def\eqnn #1\eeq{$$ #1 $$}

\def\eq #1\eeq{
\ifnum\dflag = 0
{\xdef\draftname{\ }}
\fi 
$$ #1
\eqno{\eqlabel \rlap{\ \draftname}} $$
\nexteqno}







\def\eqa #1\eeq{
\ifnum\dflag = 0
{\xdef\draftname{\ }}
\fi 
$$ \eqalignno{ #1 } $$
\global\cflag = 0}


\def\ie{{\it i.e.\/}}

\def\etc{{\it etc.\/}}
\def\etal{{\it et.al.\/}}


\def\anp#1#2#3{{\it Ann.\ Phys. (NY)} {\bf #1} (19#2) #3}

\def\jetpl#1#2#3#4#5#6{{\it Pis'ma Zh.\ Eksp.\ Teor.\ Fiz.} {\bf #1} (19#2) #3
[{\it JETP Lett.} {\bf #4} (19#5) #6]}
\def\jpa#1#2#3{{\it J.\ Phys.} {\bf A#1} (19#2) #3}

\def\npb#1#2#3{{\it Nucl.\ Phys.} {\bf B#1} (19#2) #3}
\def\plb#1#2#3{{\it Phys.\ Lett.} {\bf #1B} (19#2) #3}

\def\prb#1#2#3{{\it Phys.\ Rev.} {\bf B#1} (19#2) #3}

\def\prl#1#2#3{{\it Phys.\ Rev.\ Lett.} {\bf #1} (19#2) #3}

\def\sjnp#1#2#3#4#5#6{{\it Yad.\ Fiz.} {\bf #1} (19#2) #3
[{\it Sov.\ J.\ Nucl.\ Phys.} {\bf #4} (19#5) #6]}


\global\nulldelimiterspace = 0pt



\def\frac#1#2{{{#1} \over {#2}}\,}  
\def\hf{{1\over 2}}



\def\Asl{\hbox{/\kern-.7500em\it A}} 
\def\Dsl{\hbox{/\kern-.6700em\it D}} 
\def\dsl{\hbox{/\kern-.5300em$\partial$}}
\def\pxpsl{\hbox{/\kern-.5600em$p$}}
\def\sslsh{\hbox{/\kern-.5300em$s$}}
\def\epssl{\hbox{/\kern-.5100em$\epsilon$}}
\def\delsl{\hbox{/\kern-.6300em$\nabla$}}
\def\lxpsl{\hbox{/\kern-.4300em$l$}}
\def\elxpsl{\hbox{/\kern-.4500em$\ell$}}
\def\kxpsl{\hbox{/\kern-.5100em$k$}}
\def\qxpsl{\hbox{/\kern-.5000em$q$}}
\def\sla#1{\raise.15ex\hbox{$/$}\kern-.57em #1}



\def\roughly#1{\mathrel{\raise.3ex\hbox{$#1$\kern-.75em\lower1ex\hbox{$\sim$}}}}

\def\tw#1{\tilde{#1}}
\def\ol#1{\overline{#1}}





\def\Scl{{\cal L}}

\def\Sco{{\cal O}}


\def\sse{{\sss E}}

\def\ssh{{\sss H}}

\def\ssr{{\sss R}}

\def\sst{{\sss T}}


\def\det{\mathop{\rm det}}

\def\Re{{\rm Re\;}}
\def\Im{{\rm Im\;}}









\nopageonenumber
\baselineskip = 18pt
\barsoff

\def\bk{\item{}}

\def\qbar{{\overline{q}}}
\def\zbar{{\overline{z}}}

\def\Vbar{{\overline{V}}}
\def\plbar{{\overline{\partial}}}
\def\sxx{\sigma_{xx}}
\def\sxy{\sigma_{xy}}
\def\IR{\relax{\rm I\kern-.18em R}}
\font\cmss=cmss10 \font\cmsss=cmss10 at 7pt
\def\IZ{\relax\ifmmode\mathchoice
{\hbox{\cmss Z\kern-.4em Z}}{\hbox{\cmss Z\kern-.4em Z}}
{\lower.9pt\hbox{\cmsss Z\kern-.4em Z}}
{\lower1.2pt\hbox{\cmsss Z\kern-.4em Z}}\else{\cmss Z\kern-.4em Z}\fi}

\def\sltwoz{SL(2,\IZ)}

\def\GT{\Gamma_\sst(2)}
\def\GH{\Gamma_\ssh}
\def\HtwoT{{H_2^\sst}}
\def\EtwoT{{E_2^\sst}}


\rightline{November, 1996.}
\line{cond-mat/9611118 \hfil McGill-96/39, OSLO-TP 12-96}

\vskip .1in
\title
\centerline{One-Dimensional Flows in} 
\centerline{the Quantum Hall System}
\endtitle

\vskip 0.2in
\authors
\centerline{C.P. Burgess${}^a$ and C.A. L\"utken${}^b$}  
\vskip .1in
\centerline{\it ${}^a$ Physics Department, McGill University}
\centerline{\it 3600 University St., Montr\'eal, Qu\'ebec, Canada, H3A 2T8.}
\vskip .05in
\centerline{\it ${}^b$ Physics Department, University of Oslo}
\centerline{\it P.O. Box 1048, Blindern, N-0316 Norway.}
\endauthors

\footnote{}{\eightrm * Research supported in part by N.S.E.R.C.\ of Canada,
F.C.A.R. du Qu\'ebec and the Norwegian Research Council.}

\abstract
\vbox{\baselineskip 15pt 
We construct the $c$ function whose gradient determines the RG flow of the
conductivities ($\sigma_{xy}$ and $\sigma_{xx}$) for a quantum Hall system,
subject to two assumptions. (1) We take the flow to be invariant with respect to
the infinite discrete symmetry group, $\GH$, recently proposed by several
workers to explain the `superuniversality' of the delocalization exponents in
these systems. (2) We also suppose the flow to be `quasi-holomorphic' (which we
make precise) in the sense that it is as close as possible to a one-dimensional
flow in the complex parameter $\sigma_{xy} +i \sigma_{xx}$. These assumptions
together with the known asymptotic behaviour for large $\sigma_{xx}$, completely
determine the $c$ function, and so the phase diagram, for these systems. A
complete description of the RG flow also requires a metric in addition to the
$c$ function, and we identify the features which are required for this by the
RG. A similar construction produces the $c$ function for other systems enjoying
an infinite discrete symmetry, such as for supersymmetric QED. }     
\endabstract


\vfill\eject

\ref\pruiskenbeta{E. Br\'ezin, S. Hikami and J. Zinn-Justin,
\npb{165}{80}{528}; \bk  
S. Hikami, \plb{98}{81}{208}; \bk
A. Pruisken, \prb{31}{85}{416}.}

In this article we present a conjecture concerning the exact beta function
describing the renormalization group (RG) flow of the conductivities for
quantum Hall systems. We do so by deriving the most general possible $c$
function which is consistent with three properties which we assume for this
flow. (The RG beta function may be obtained from the $c$ function by taking its
gradient.) The properties which we use are: ($i$) the flow commutes with an
infinite discrete symmetry group, $\GH$; ($ii$) it has a beta function with no
singularities and which is consistent with the large-$\sxx$ form as predicted
by the weak-coupling expansion of the effective sigma-model description of the
RG \pruiskenbeta; and ($iii$) the flow is `quasi-holomorphic' in the complex
conductivity, $z$, defined by: $\sigma_{xy}+i\sigma_{xx} = \left({e^2 \over h}
\right)z$. These assumptions are sufficiently strong to completely determine the
$c$ function, and so permit the identification of the phase diagram in both the
strong-coupling (small $\sxx$) and weak-coupling (large $\sxx$) regimes. 

\section{The Assumptions} 

Before describing our results, some justification for the three requirements
are in order. We deal with each in turn. 

\topic{(1) The Symmetry}
Convincing experimental evidence and theoretical arguments support the existence
of an infinite discrete symmetry, $\GH$, of the parameter space in the quantum
Hall system. There are three main lines of evidence.

\ref\scalingexp{H.P. Wei, D.C. Tsui, M.A. Paalanen and A.M.M. Pruisken,
\prl{61}{88}{1294}; \bk
L. Engel, H.P. Wei, D.C. Tsui and M. Shayegan, {\it Surf. Sci.} {\bf 229}
(1990) 13; \bk
S. Koch, R.J. Haug, K.v. Klitzing and K. Ploog, \prl{67}{91}{883}.}

\ref\numscaling{J.T. Chalker and P.D. Coddington, {\it J.\ Phys.} {\bf C21}
(1988) 2665; \bk
B. Huckestein and B. Kramer, \prl{64}{90}{1437}.}

\ref\lutkenross{C.A. L\"utken and G.G. Ross, \prb{45}{92}{11837}; 
\prb{48}{93}{2500}; \bk 
C.A.L\"utken, \jpa{26}{93}{L811}; \npb{396}{93}{670}.}

First, the delocalization exponent which describes the transition between two
quantum Hall phases, as found both in real scaling experiments  \scalingexp\ and
numerical simulations \numscaling, exhibits a ``superuniversality'' in the sense
that it does not depend on the which  phases are involved. This otherwise
mysterious equivalence is easily understood as a consequence of a discrete
symmetry, $\GH$, mapping the critical points into one other.\foot\append{See
Appendix A for this argument in detail.}  

Furthermore, the positions of these critical points are consistent with what
would be expected if $\GH$ were a large subgroup of the modular group, 
$\GH \subseteq \sltwoz$, where the modular group is defined to act on the 
complex conductivity, $z$, by holomorphic fractional linear 
transformations \lutkenross:
\label\sltwozdef
\eq
z \to \gamma (z) = \frac{az + b}{cz + d},\quad ad-bc =1,\quad
a,b,c,d \in \IZ.
\eeq
In fact, the very existence of the quantum Hall plateaux at  fractional values
of  $\sigma_{xy}$ is an inevitable and striking consequence of such a symmetry 
if $\GH$ is a sufficiently large subgroup of the modular group.

\ref\fradkin{E. Fradkin, {\it Field Theories of Condensed Matter Systems},
Addison Wesley, 1991.}

\ref\BLQ{C.P. Burgess, C.A. L\"utken and F. Quevedo, \plb{336}{94}{18},
hep-th/9407078; \bk
J. Fr\"olich, R. G\"otschmann and P.A. Marchetti, 
\jpa{28}{95}{1165}, hep-th/9406154.}

\ref\calkids{S.-C. Zhang, T.H. Hansson and S. Kivelson, \prl{62}{89}{82}; \bk 
S.-C. Zhang, {\it Int.\ J.\ Mod.\ Phys.} {\bf B6} (1992) 25.}

\ref\klz{S. Kivelson, D.-H. Lee and S.-C. Zhang; \prb{46}{92}{2223}.}

Secondly, such a symmetry is also very plausible on theoretical grounds. 
On the Hall plateaux the long wavelength electromagnetic response is described
by a Chern-Simons action, $\Scl_{\sss CS} = - \hf \, \sxy \,
\epsilon^{\mu\nu\lambda} A_\mu \partial_\nu A_\lambda + e j^\mu A_\mu$. But
general arguments \fradkin\ imply that such an action is invariant under the
replacement $\sxy^{-1} \to \sxy^{-1} + 2n \,{h\over e^2}$, where $n$ is any
integer. Similar arguments as applied to the dual action \BLQ\ imply a symmetry
with respect to shifts of $\sxy$ rather than its inverse.  Furthermore, a
convincing picture of the full symmetry group also emerges from more
microscopic considerations, such as expressed by the ``Law of Corresponding
States'' as formulated within the context of the Chern-Simons Landau-Ginzburg
(CSLG) theory of the quantum Hall effect \calkids, \klz. 

\ref\ivcharacter{D. Shahar \etal, Princeton preprint, cond-mat/9510113.}

Finally, a remarkable recent experiment \ivcharacter\ has shown that 
quantum Hall systems in different phases, even far from the transition
point, have nonlinear $I-V$ response curves which are dual to one 
another in a well-defined way that is consistent 
with the particle-vortex duality that one expects from the CSLG picture.
Remarkably, the persistence of duality far from the transition point 
seems to indicate that its action is not destroyed by renormalization effects.  

\ref\dualguys{L. P. Pryadko and S.-C. Zhang, Stanford preprint SU-ITP \#95/27,
cond-mat/9511140; \bk
E. Fradkin and S. Kivelson, cond-mat/9603156.}

All of these considerations point to the symmetry group being the subgroup
$\GH = \GT \subset \sltwoz$, which is generated by the two transformations
\lutkenross:\foot\otherduals{Duality and discrete symmetries in condensed matter
systems have recently been treated from a different point of view in 
ref.~\dualguys.}   
\label\gammaTgens
\eq
T(z) = z + 1\qquad {\rm and}\qquad R(z) = \frac{z}{1 - 2z}. 
\eeq

\topic{(2) The Large-$\sxx$ Limit}
The transport properties for quantum Hall systems have been argued to be 
described in the far infrared by a two-dimensional sigma model. The beta
function for the complex conductivity, $z$, may be evaluated in this  model for
large $y = \Im z$ using semiclassical methods. In perturbation theory the result
is independent of $x = \Re z$ and is given by \pruiskenbeta: 
\label\sigmamodelresult
\eq
\left( {dz \over dt} \right)_{\rm p.t.} \equiv \beta^z(x,y)_{\rm p.t.} 
= -\, {i \over 2 \pi^2 y} +O  \left({1 \over y^2}\right).
\eeq
The real part, $x$, first enters the evolution equations in the large-$y$ limit
through nonperturbative instanton contributions, which contribute to $\beta^z$ a
series in the quantity $\qbar = e^{- 2 \pi (y + ix)}$, with each term of this
series premultiplied by its own series in $1/y$. We therefore demand a
similar large-$y$ limit for $\beta^z$ --- \ie\ a Taylor series in $q$ and
$\qbar$ whose coefficients might themselves be a series in $1/y$. For later
purposes notice that once the perturbative series in $1/y$ is put aside, the
instanton contributions to ${dz \over dt}$ depend on the {\it anti}-holomorphic
quantity $\zbar = x-iy$. 

\topic{(3) Quasi-Holomorphy}
If $\beta^z$ were a holomorphic function --- that is, independent of $\zbar$ ---
then powerful results of complex analysis could be used to rule out the
existence of a beta function satisfying the previous two assumptions. 
Even if this were not so, the asymptotic expansion in powers of $1/y$ explicitly
breaks holomorphy and so excludes a holomorphic $\beta^z$. The assumption of
`quasi-holomorphy' is meant to capture the predictive nature of holomorphy in a
way which is consistent with the asymptotic form of the sigma model at large
$y$. We postpone its detailed definition until further tools are introduced, and
simply motivate it here as a very predictive {\it ansatz} which is motivated by
the asymptotic expression, eq.~\sigmamodelresult. It is also suggested by the
experimental observation \ivcharacter\ of duality far from the critical points,
since these seem to indicate that the holomorphic group action of
eq.~\sltwozdef\ is not corrupted by renormalization effects. It is hard to
imagine how this would be possible if the complex conductivities $z$ and $\zbar$
were to mix under rescalings of the system. 

In any case, quasi-holomorphy incorporates the physical requirements of the
system in the simplest possible way. If its predictions are borne out,
then we will acquire an important constraint on  microscopic descriptions of
these systems.     
\endtopic

We next turn to the consequences of these assumptions, including a precise
definition of quasiholomorphy. As shall become clear, the resulting conditions
on the RG flow are very restrictive, and at present we have only been able to
solve them for the $c$ function, and not also for the delocalization exponents,
$\nu$, at the critical points. 

\section{The Consequences}

\ref\peterh{P.E. Haagensen, \plb{382}{96}{356}; \bk
P.H. Damgaard and P.E. Haagensen, Neils Bohr/MIT preprint
NBI-HE-96-50 MIT-CTP \#2569, cond-mat/9609242.}

\ref\rankin{Robert R. Rankin, {\it Modular Forms and Functions}, Cambridge
Univeristy Press, 1977; \bk
Neal Koblitz, {\it Introduction to Elliptic Curves and Modular Forms}, 2nd
Edition, Springer-Verlag, 1984}

We are therefore led to investigate the properties of nonsingular beta functions
whose flows are compatible with a sub-modular group $\Gamma$.\foot\peterref{
Ref.~\peterh\ contains similar applications of finite discrete groups to
constraining beta functions.} The basic requirement is that the action of the
symmetry, $z \to \gamma(z)$, should commute with the RG flow, $z \to
r_t(z,\zbar)$. That is (see Appendix A for details): $\gamma \circ r_t = r_t
\circ \gamma$. For infinitesimal flows this implies the beta function, $\beta^z
= {dz \over dt}$, must satisfy:  
\label\betatransfn
\eq
\beta^z(\gamma(z),\ol{\gamma(z)}) = {d\gamma \over dz} \; \beta^z(z,\zbar)
=  (cz + d)^{w} \beta^z(z,\zbar),
\eeq
with $w=-2$. For $\beta^z$ holomorphic, the second equality in eq.~\betatransfn\
defines an automorphic function of weight $w$. It is a theorem \rankin\ that no
such function having negative weight exists which is analytic throughout the
upper half plane,\foot\imzispos{Recall $\ss \Im z \propto \sxx \ge 0$.} and
which admits an expansion in integer powers of $q$ as $z \to i\infty$. No such
candidate beta function can therefore exist. 

\ref\zamolodchikov{A.B. Zamolodchikov, \jetpl{43}{86}{565}{43}{86}{730};
\sjnp{46}{87}{1819}{46}{87}{1090}.}

We must therefore relax the condition that $\beta^z$ be holomorphic. To this end
imagine using a metric\foot\metricnotes{Although we do not use any particular
metric in what follows, a natural metric for these purposes is the Zamolodchikov
metric, which is defined in terms of the  correlations of the relevant operators
of the sigma model \zamolodchikov. Since the sigma model becomes weakly coupled
in the large-$\ss y$ regime, this metric might be expected to approach the
flat Cartesian metric $\ss ds^2 = dz \, d\zbar$ in this limit.}  $G_{ij}$ to
convert the contravariant quantity, $\beta^j$, into a covariant vector:
$\beta_{z} = G_{zz}\beta^{z} + G_{z\zbar}\, \beta^{\zbar}$. Since $G_{ij}$ is
typically not holomorphic it is clear that we cannot demand that {\it both}
$\beta^z$ {\it and} $\beta_z$ be holomorphic, or almost so. 

\ref\derivflow{D.J. Wallace and R.K.P. Zia, \anp{92}{75}{142}.}

\ref\cfl{A. Cappelli, D. Friedan and J.I. Latorre, \npb{352}{91}{616}.}

We may now more precisely state the condition of quasi-holomorphy. We assume the
{\it covariant} quantity, $\beta_z$, to be a holomorphic  function of $z$ (and
not $\zbar$), apart from the addition of a possible series in $1/y$. We also
take it to be nonsingular throughout the upper half complex plane. There are
three motivations for making this {\it ansatz} for $\beta_z$ rather than for
$\beta^z$. First, recall that sigma-model calculations produce instanton
corrections to $\beta^z$ which depend on $\zbar$ instead of $z$. This might be
expected if $\beta^z$ were obtained as $G^{z\zbar} \beta_\zbar$ with
$\beta_\zbar$ antiholomorphic. Second, it is believed \derivflow,
\zamolodchikov, \cfl\ that all RG flows in two (and possibly higher) dimensions
are {\it gradient} flows, in that $\beta_i = \partial_i \Phi$ for a potential
function, $\Phi$. In this sense it is $\beta_i$ which is more fundamental, and
so which may be expected to have simpler properties. Third, the function
$\beta_z$ transforms as an automorphic function of weight $w=+2$ with respect to
$\GT$ transformations, and it is a theorem \rankin\ that this is the lowest
possible weight for which such a holomorphic function exists. 

\subsection{The Conditions Governing $\beta_i$}

The three conditions are now sufficiently precise to be used to constrain the
possible form for the beta function. For brevity's sake, we simply quote our
main results here. $\beta_z$ may be conveniently constructed in terms of the
generalized Eisenstein series: 
\label\Eisenlattice
\eq
E^{(p,q)}_k(z) \equiv N_k {\sum_{mn}}' (mz + n)^{-k},
\eeq
where the sum over $m$ and $n$ is over all pairs of integers $(m,n)$ --- the
prime indicating the omission of $(0,0)$ --- for which $(m,n) = (p,q)$ (mod 2). 
The normalization constant, $N_k$, is chosen to ensure $E_k^{(p,q)} \to 1$ as
$y\to \infty$, and is given in terms of the Riemann zeta function,
$\zeta_\ssr(k)$, by $N_k^{-1} = 2 \zeta_\ssr(k) \, (1 - 2^{-k})$. The
series converges for $k \ge 2$, and is analytic throughout the upper half plane.
For $k > 2$ the sum defines an automorphic function of weight $w=k$. The unique
(up to normalization) $w=2$ holomorphic function for the group $\GT$ is given by
\rankin: 
\label\holotwo
\eq
\EtwoT(z) \equiv - \, {3\over 8}\; E^{(0,0)}_2(z) + {3 \over 2} \;
E^{(0,1)}_2(z)  = 1 + 12\sum_{n=1}^{\infty} [\sigma_1^+(n) - \sigma_1^-(n)]
\;q^n , 
\eeq
where $q = e^{2\pi i z}$ and $\sigma_k^{\pm}(n) = \sum_{d\vert n} (\pm)^d d^k $
is a sum over all positive integers $d$ that divide $n$. A second weight-two
function, which is {\it quasi}holomorphic, may be chosen to be:
\label\nonholotwo
\eq
\HtwoT(z,\zbar) \equiv {1 \over \pi y} - \, {3\over 8}\; E^{(0,0)}_2(z) + 
{1 \over 2} \; E^{(0,1)}_2(z)  = \frac{1}{\pi y}
+ 4\sum_{n=1}^{\infty} [3\sigma_1^+(n) - \sigma_1^-(n)] \;q^n .  
\eeq
\ref\hecke{E. Hecke, {\it Abh. Math. Sem. Univ. Hamburg.}, {\bf 5} (1927) 199.}
We use $H$ to denote such quasiholomorphic quantities in recognition of Hecke,
who first introduced them \hecke. These are the only independent linear
combinations of the $E^{(p,q)}_2(z)$ which transform with weight two after the
addition of a function of $y$, and we assume these to form a basis for all
quasiholomorphic functions having weight two.

If we use eq.~\sigmamodelresult\ for the asymptotic form for large $y$, and 
if we suppose $G_{z\zbar} = 1 + O(1/y)$ and $G_{zz} = O(1/y)$ in this limit,
then we must choose:  
\label\covresult
\eq
\beta_z ={i \over 2 \pi} \, \HtwoT(z,\zbar).
\eeq
If, on the other hand, we permit $\beta_z = O(1)$ for large $y$, then it may be
taken to be an arbitary linear combination of $\HtwoT(z,\zbar)$ and $\EtwoT(z)$.

Both $\EtwoT$ and $\HtwoT$ are nonsingular in the upper half plane, and have
simple zeroes at the fixed point $\xi \equiv \hf \,(1 + i)$ of $\Gamma_T(2)$,  
as well as at all its images under the group. Since $\det G_{ij}$ should never
vanish, these are therefore the candidates for the delocalization critical
point in the  quantum Hall system. We obtain in this way the phase diagram of
refs.~\lutkenross\ and \klz. Since these critical points are all related to one
another by $\GT$ transformations, they must all share the same critical
exponents. The values for these exponents are found by expanding $\beta^i$ about
$z=\xi$, and computing the inverses of the eigenvalues of the following matrix
of derivatives: 
\label\critexps
\eq
(\partial_i \beta^j) \Bigr|_{z=\xi} = G^{jk}(\xi) \; (\partial_i
\beta_k) \Bigr|_{z=\xi}. 
\eeq
Clearly a determination of these indices requires knowledge of the inverse
metric evaluated at the critical point. 

It is noteworthy that choosing $\beta_z$ to be any linear combination of
$\HtwoT$ and $\EtwoT$ gives a gradient flow, $\beta_z = \partial
\Phi$.\foot\derivs{Here, and in what follows, $\ss \partial$ without a subscript 
denotes $\ss \partial/\partial z$, and $\ss \plbar$ denotes $\ss
\partial/\partial \zbar$.} This is because both $\HtwoT$ and $\EtwoT$ are the
derivatives of appropriate functions: $i\HtwoT = \partial \Phi_\ssh$ and $i
\EtwoT = \partial \Phi_\sse$,  with:    
\label\RGPotresult
\eq
\Phi_\ssh = {1 \over 2 \pi} \, \ln \left| \frac{\Delta_\sst}{y^4 \Delta} 
\right| , \qquad \hbox{and} \qquad  
\Phi_\sse = {1 \over 2 \pi} \, \ln \left| { \Delta_\sst^3 \over
\Delta} \right|  . 
\eeq
Here $\Delta$ and $\Delta_\sst$ are the so-called `cusp' forms of lowest weight
for $\sltwoz$ and $\GT$ which are analytic everywhere, and vanish nowhere, in
the upper half plane (excluding $i\infty$). They are given in terms of the
Eisenstein series by:\foot\norms{We adopt here an unconventional normalization
for $\ss \Delta$.} 
\label\Deltadefs
\eq
\Delta(z)   = (E_4)^3 - (E_6)^2,\qquad
\Delta_\sst(z) = E_4 - E_4^{(0,1)} .
\eeq
These expressions use the $\sltwoz$ Eisenstein series, $E_k(z)$, which are
defined in terms of eq.~\Eisenlattice\ by $E_k(z) = (1 - 2^{-k}) \,
E_k^{(0,0)}(z)$. 

\subsection{The Conditions Governing $G_{ij}$}

Obtaining the RG flow given $\beta_z$ requires raising the index using a
metric. We now consider what requirements govern its possible form. We
formulate two such conditions, which are required to ensure consistency with
the assumed form for $\beta_i$. We require:

\topic{(1) $\GT$ Covariance} The symmetry, $\GT$, acts on the coupling constants
as indicated in eq.~\sltwozdef. $G_{ij}(z,\zbar)$ must transform under
this transformation as a covariant rank-two tensor. We therefore require the
following transformation property under $\GT$: 
\label\metrictransfn
\eq
G_{zz} \to (cz +d)^4 \; G_{zz}, \qquad \hbox{and} \qquad 
G_{z\zbar} \to |cz +d|^4 \; G_{z\zbar} .
\eeq

\topic{(2) RG Consistency} There is a consistency requirement that the metric
must satisfy which arises because the quasiholomorphy assumption is imposed on
the components of $\beta_i$ rather than on $\beta^i$. To understand the issue,
imagine examining $\beta_i$ at two points, $z$ and $z'$, which are related by an
RG transformation: $z' = r_t(z,\zbar)$. There are then two ways to compute the
components of $\beta_i$. First, one can simply evaluate the ansatz for $\beta_i$
at both $z$ and $z'$. Second, one can compute $\beta_i'(z',\zbar')$ by evolving
$\beta_i(z,\zbar)$ from $z$ to $z'$ using the RG evolution. These two
alternatives need not agree with one another given an arbitrary ansatz for the
metric, $G_{ij}(z,\zbar)$. We must require the metric to be consistent, in this
sense, with our ansatz for $\beta_z$. 

To formulate this condition more precisely, suppose we adopt an ansatz for the
metric:
\label\metricansatz
\eq
G_{ij}(z,\zbar) = \sum_a c_a \, G^{(a)}_{ij}(z,\zbar),
\eeq
where $c_a$ are constants and each of the $G^{(a)}_{ij}$'s is a tensor having a
specific functional form. For example, we might choose one of these to be the
hyperbolic metric: $G^{(1)}_{ij} \, dz^i dz^j = (2/y^2) \, dz d\zbar$, or we
could choose $G^{(2)}_{ij} \, dz^i dz^j = V^2 \, dz^2 + \Vbar^2 \, d\zbar^2 + 2
V\, \Vbar \, dz \, d\zbar$, \etc, with $V \equiv \HtwoT$, say. (Notice that both
of these examples transform properly with respect to $\GT$.) The consistency
condition states that the change of $G_{ij}$ due to the RG flow must also have
the same form as eq.~\metricansatz. That is:
\label\dmetriccheck
\eq
\eqalign{
(\Scl_\beta G)_{ij} &\equiv \beta^k \, \partial_k G_{ij} + G_{ik} \,
\partial_j \beta^k + G_{kj} \, \partial_i \beta^k \cr
&\equiv \nabla_i \beta_j + \nabla_j \beta_i \cr 
&= \sum_a \tw{c}_a \, G^{(a)}_{ij}(z,\zbar),\cr} 
\eeq
where $\Scl_\beta G$ is defined by the first line of eqs.~\dmetriccheck,
and denotes the Lie derivative of $G_{ij}$ along the RG flow. The second
equality in eq.~\dmetriccheck\ is an identity, with the covariant derivative,
$\nabla_i$, constructed using the Levi-Civita connection for the metric
$G_{ij}$. The content of the consistency condition lies in the third of
eqs.~\dmetriccheck. When it is satisfied, the RG evolution simply makes the
parameters of ansatz \metricansatz\ into functions of scale: $c_a = c_a(t)$. 

\endtopic

To date we have been unsuccessful in finding a simple ansatz for $G_{ij}$ which
satisfies condition \dmetriccheck. It would be encouraging to think that this
condition may be sufficiently restrictive as to ensure a unique solution. 

\section{The \sltwoz\ Case}

We pause here to briefly record the results which may be obtained using
identical arguments for the case where the symmetry group is $\sltwoz$ itself,
rather than just its subgroup, $\GT$. These results may have applications
within supersymmetric gauge theories, such as for supersymmetric QED with both
light elementary charges and magnetic monopoles. 

Only one linear combination of the two forms, $\HtwoT(z,\zbar)$ and
$\EtwoT(z)$, transforms with weight $w=2$ under the full group, $\sltwoz$. This
unique combination is given by:
\label\sltwozetwo
\eq
H_2(z,\zbar) = - \; {3 \over \pi y} + E_2(z) = 1 - \; {3 \over \pi y} 
- 24 \sum_{n=1}^\infty \sigma^+_1(n) \, q^n,
\eeq
where, $E_2 = {3 \over 4} \, E_2^{(0,0)}$. Such a covariant beta function may
be written as the derivative of a real RG potential in the following way:
$i H_2(z,\zbar) = \partial \Psi(z,\zbar)$, with: 
\label\sltwozrgpotential
\eq
\Psi = { 1 \over 2 \pi} \; \ln \Bigl( y^{12} \, \Delta \, \overline{\Delta}
\Bigr). 
\eeq

The beta function defined by eq.~\sltwozetwo\ has two types of fixed points,
corresponding to the points $z = i$ and $z = e^{i \pi/3}$, as well as the
images of these under $\sltwoz$. The resulting phase diagram is described in
more detail in ref.~\lutkenross. 

\section{Conclusions}

In this paper we have argued that plausible features of quantum Hall systems,
together with the ansatz of quasiholomorphy, combine to strongly constrain the
RG flow which is possible for these systems. We have used these constraints to
determine the $c$ function, from which the beta function is obtained by taking
the gradient. We found a two-parameter space of solutions, which can be
further restricted by using the weak-coupling form which has been computed
for quantum Hall systems for large $\sxx$. For asymptotically flat metrics,
the result is unique, and is given by eq.~\RGPotresult:
\eq
\beta_z = {i \over 2\pi} \; \HtwoT(z,\zbar) = \partial \Phi \qquad
\hbox{with} \qquad \Phi = {1 \over 2 \pi} \; \Phi_\ssh = \left({1 \over 2
\pi}\right)^2 \, \ln \left| \frac{\Delta_\sst}{y^4 \Delta} \right| . 
\eeq
This function fixes the phase diagram for these systems even in the strong
coupling (small $\sxx$) regime. The extraction of more information requires
knowledge of a Zamolodchikov-type metric, which relates $\beta^i$ to $\beta_j =
\partial_j \Phi$. 

We close by listing some of the many directions which remain to be pursued.
First, an explicit solution of the consistency condition, eq.~\dmetriccheck,
for $G_{ij}$ would be useful for quantum Hall systems, since this would permit
the calculation of the delocalization critical exponents, $\nu$. Should such a
prediction succeed, it would immediately suggest searching for the underlying
origin of the symmetry and quasiholomorphy properties. It is perhaps
tantalizing in this regard that disordered systems such as these are often
describable using supersymmetric models, since supersymmetric models often
exhibit special holomorphy properties. 

\appendix{A}{Beta functions and Discrete Symmetries}

We start by reviewing properties shared by all beta functions.

\subsection{Definitions and Notation}

Consider a system which for some range of energies is adequately
described by an effective Lagrangian parametrized by $n$ independent
couplings $\{g^1,g^2,\dots,g^n\}$.  Scale transformations on the
system generates a flow in this parameter space:
\label\flow
\eq
g^i \rightarrow r_t^i(g) = g^i(t)
\eeq
labelled by the scale parameter $t = \ln (\mu/m)$, where $\mu$ is the
mass scale of interest and $m$ is a fiducial mass scale relative to
which we choose to measure $\mu$.  The beta function is a vector field
tangent to this flow:
\eq
\beta^i(g) = \frac{dg^i}{dt}.
\eeq

\ref\netricdef{D. O'Connor and C.R. Stephens, in the proceedings of the 
1993 International Symposium (Maryland), Vol 1, eds. B.L. Hu, M.P. Ryan Jr. and
C.V. Vishevshawara, Cambridge University Press, 1993 (hep-th/9304095).}

Notice that $\beta^i$ transforms like a contravariant vector field under
general coordinate transformations on the parameter space; i.e. if
$g^i\rightarrow g^{\prime i}$ then
\label\betatransform
\eq
\beta^i \rightarrow \beta^{\prime i} 
= \frac{\partial g^{\prime i}}{\partial g^j} \beta^j
=  g^{\prime i}_{~,j} \beta^j . 
\eeq
The covariant vector field  $\beta_i$ is obtained from $\beta^i$ by lowering
the index with a metric $G_{ij}$ on the parameter space, $\beta_i =
G_{ij}\beta^j$. In a specific quantum field theory, such a metric can be
obtained \zamolodchikov, \netricdef\ using the two-point correlation functions of
the operators $\Sco_i$ to which the $g^i$ couple. That is, if the interaction
lagrangian is $\Scl = g^i \, \Sco_i$, then $G_{ij} \simeq \int dx \;
\langle\Sco_i(x) \Sco_j(0)\rangle$. 

Since the RG flows physically express how the system responds as successive 
degrees of freedom are `integrated out', they are believed to be always
monotonic, i.e. flowlines flow from sources to sinks, and never close apon
themselves. This behaviour is automatic if the flow is a gradient flow: 
\eq
\beta^i = G^{ij} \partial_j\Phi,
\eeq
since in this case $\beta^i$ is curl-free. The function $\Phi$ is known as the
RG potential. For two-dimensional, unitary field theories, all RG flows are
known to be gradient in the immediate vicinity of a fixed point \zamolodchikov.
Moreover, they have been found to be gradient in all other unitary field
theories that have been checked. 

\subsection{Behaviour Near Fixed Points}

$g = g_*$ is a fixed point of the RG flow if the beta function vanishes there.
Expanding around such a point, 
\eq 
\beta^i(g) = B^i_{~j}(g - g_*)^j + O((g-g_*)^2)
\eeq
we obtain a differential equation
\eq
\frac{dg}{dt} = B^i_{~j}(g - g_*)^j
\eeq
whose solution is
\eq
(g - g_*)^i = ( e^{B(t - t_0)} )^i_{~j}(g - g_*)^j,
\eeq
where $g_0 = g(t_0)$ is the starting point (initial value) of the
flow.  Rotating to the basis $\tilde g^i$ where $B$ is diagonal,
\eq
(\tilde g - \tilde g_*)^i = e^{b_i(t - t_0)}(\tilde g - \tilde g_*)^i,
\eeq
we see that the RG flow decouples into one-dimensional flows near the
fixed points, and that the eigenvalues of $B$ give the flow rates
in the principal directions around the fixed point.

In order to see how the eigenvalues $b_i$ are measured, we consider
any physical quantity $\xi$ defined such that it depends on the scale
transformation
\eq
\mu=\lambda\mu, \quad t = t_0 + \ln \lambda,\quad \lambda(t) =
e^{t-t_0}
\eeq
only through $g$:
\eq
\xi(g_0) = \xi (g(t_0))\rightarrow \xi(g(t)) =
\lambda^{-1}(t)\xi(g_0).
\eeq
Differentiating with respect to $t$ and letting $t\rightarrow t_0$ we obtain a
differential equation for $\xi$ ($\xi_{,i} \beta^i + \xi = 0$) which to linear
order in $g$ is 
\eq
\xi_{,i}B^i_{~j}(g - g_*)^j + \xi = 0.
\eeq
In the basis which diagonalizes $B$ the solution separates
\eq
\frac{\xi}{\xi_0} = \prod_{i=1}^n \frac{\xi^i}{\xi^i_0}
\eeq
where the $\xi^i = (\tilde g^i - \tilde g^i_*)^{-\nu_i}$ can be regarded as
``decoupled'' correlations lengths which  define the observable critical scaling
exponents $\nu_i = b^{-1}_i$.  Any physical length scale $l$ can be expressed in
terms of the $\xi^i$ through a homogenous function $F$ of weight one:
\eq
l = F(\xi^1,\xi^2,\dots,\xi^n) = \lambda^{-1}
F(\lambda\xi^1,\lambda\xi^2,\dots,\lambda\xi^n).
\eeq
Equivalently, with  quasi-homogenous parameters $x^i = (\tilde g -\tilde g_*)^i$
of quasi-weight $-\nu_i^{-1}$, $l$ is a quasi-homogeneous function $f$ of total
weight one: 
\eq
l = f(x^1,x^2,\dots,x^n) = \lambda^{-1}
f(\lambda^{-\frac{1}{\nu_1}}x^1,\lambda^{-\frac{1}{\nu_2}}x^2,\dots,
\lambda^{-\frac{1}{\nu_n}}x^n) .
\eeq
Universality is the idea that all lagrangians with a given symmetry flow to the
same infrared (IR: $t\rightarrow +\infty$, UV: $t\rightarrow -\infty$) fixed
point.  In physical systems exhibiting second order phase transitions, the
critical points are RG fixed points with unstable ($\lambda_i > 0$) directions. 
It is an experimental fact that the critical exponents $\nu_i$ measured in such
systems take only a few specific values, which therefore label the
corresponding universality classes.  In short, different theories can
and do have the same critical behaviour.

\subsection{RG and Discrete Symmetries}

In the body of this paper we are concerned with a different type of
universality, sometimes called ``superuniversality'', where different fixed
points have the same critical behaviour.  In particular, {\it distinct critical
points have identical scaling exponents}. We next show in detail why 
such behaviour is to be expected when the system possesses a set of {\it
parameter space} symmetries $g^i\rightarrow \gamma^i(g)$ which commute with the
RG flow: 
\eq
\gamma(r_t(g)) = r_t(\gamma(g)).
\eeq
In order to see this, differentiate the equation with respect to 
$t$ and evaluate at $t=0$, giving:
\label\eigeneq
\eq
\gamma^i_{~,j}(g) \, \beta^j(g) = \beta^i(\gamma(g))
\eeq
for all $g$.   This is just eq. \betatransform ~specialized to
the symmetry transformations, $g^{\prime i}(g) = \gamma^i(g)$.
Differentiating again gives
\label\fp
\eq
\gamma^i_{~,jk}(g)\beta^j(g) + \gamma^i_{~,j}(g)\beta^j_{~,k}(g) =
\beta^i_{~,j}(\gamma(g)) \, \gamma^j_{~,k}(g).
\eeq
If two distinct RG fixed points $g_*$ and $g_{**}$ are related by
$\gamma$ then
\eq
g^i_{**} = \gamma^i(g_*)\quad {\rm and} \quad \beta^i(g_*) = \beta^i(g_{**})
= 0,
\eeq
and  eq. \fp ~reduces to
\label\fpfp
\eq
\gamma^i_{~,j}(g_*)\beta^j_{~,k}(g_*)=\beta^i_{~,j}(g_{**})\gamma^j_{~,k}(g_*).
\eeq
Finally, expanding $\beta(g)$ around $g_*$ and $g_{**}$ with linear
matrix coefficients $B_*$ and $B_{**}$, respectively, and expanding
$\gamma(g)$ around $g_*$,
\eq
\gamma^i(g) = g^i_{**} + D^i_{~j}(g - g_*)^j + \dots
\eeq
we find from eq. \fpfp ~that $B_{**} = D B_* D^{-1}$.  Since $B_{**}$ is simply
a  similarity transformation of $B_*$ they must have the same eigenvalues,
i.e. the same critical exponents.  

Notice also that if $g_*$ is fixed by some transformation $\gamma_*$
in the symmetry group $\Gamma$, then eq. \eigeneq ~shows that
either $\gamma^i_{*,j}$ has $1$ as an eigenvalue, or $\beta(g_*) = 0$ 
and $g_*$ must also be an RG fixed point.   The former case arises
below when we turn to modular transformations of the upper half of the
complex plane. The point $z_* = i\infty$, which will
play a central role in our discussion, is fixed by translations 
$T(z) = z + 1$, but $z_*$ need not be an RG fixed point and the beta
function need not vanish at this point.  Other conditions on the beta
function, like holomorphicity, may force it to vanish, also at
$i\infty$, but it is important to realize that it is not sufficient 
that $z_*$ is a fixed point of $\Gamma$ for it to be a fixed point of 
the RG flow. 

\bigskip
\centerline{\bf Acknowledgments}
\bigskip

Many aspects of this work have benefited from fruitful discussions with
Dan Arovas, Peter Haagensen, Rob Myers, Fernando Quevedo and Shou-Cheng Zhang.
We are grateful to the organizers of the Oslo workshop  on 
low-dimensional physics (June, 1996), where some of this work was done.
This research was partially funded by the N.S.E.R.C.\ of Canada  and the
Norwegian Research Council.

\listrefs

\bye